\documentclass[preprint,aps,showpacs]{revtex4}
\usepackage{graphicx}

\def\be7pg{$^7Be(p,\gamma)^8B$}
\def\xbe7{$^7Be$}
\def\b8{$^8B$}
\def\S17{$S_{17}(0)$}
\def\xs17{$S_{17}$}
\def\s34{$S_{34}(0)$}
\def\xpm{$\pm$}

\begin{document}

\title{Status of the Standard Solar Model 
Prediction of Solar Neutrino Fluxes}
\thanks{Work Supported by USDOE Grant No. DE-FG02-94ER40870.}

\author{Moshe Gai}
\affiliation{Laboratory for Nuclear Science at Avery Point,
University of Connecticut, 1084 Shennecossett Rd., Groton, CT 06340-6097, USA \\  
e-mail: moshe.gai@yale.edu, URL: http://astro.uconn.edu}

\begin{abstract}
The Standard Solar Model (BP04) predicts a total \b8 neutrino flux that is 17.2\% larger 
than measured in the salt phase of the SNO detector (and if it were significant it will 
indicate oscillation to sterile neutrinos). Hence it is important 
to examine in details uncertainties (and values) of inputs to the SSM.
Currently, the largest fractional uncertainty is due to the new evaluation of the 
surface composition of the sun. We examine the nuclear input on  
the formation of solar \b8 [\S17] and demonstrate that it is still quite uncertain due to 
ill known slope of the measured astrophysical cross section factor and thus ill 
defined extrapolation to zero energy. This yields an additional reasonably estimated 
uncertainty due to extrapolation of $^{+0.0}_{-3.0}$ eV-b ($^{+0\%}_{-14\%}$).  
Since a large discrepancy exists among  
measured as well as among predicted slopes, the value of \S17 is dependent 
on the choice of data and theory used to extrapolate \S17. This situation must 
be alleviated by new measurement(s). The "world average" 
is driven by the Seattle result due to the very small quoted uncertainty, which we 
however demonstrate it to be an over-estimated accuracy. We propose more realistic 
error bars for the Seattle results based on the published Seattle data.

\end{abstract}


\maketitle

\section{Introduction} 

The high precision measurement of neutral current interactions of "\b8 
solar neutrinos" in the SNO detector (with added salt) \cite{SNO}, yields the measured flux: 
$\Phi_{NC}$ = 4.94 \xpm \ 0.21 \ (stat) $^{+0.38}_{-0.34}$ \ (syst) x $10^6$ cm$^{-2}$ sec$^{-1}$, 
with a total uncertainty of +8.8 -8.1\%. The most realistic Standard Solar Model (SSM) 
prediction for the \b8 flux, 
labeled as BP(04) in \cite{BP04}, is 5.79 x $10^6$ cm$^{-2}$ sec$^{-1}$, hence: 

\begin{center}
${\Large{\Phi_{SSM} \over \Phi_{SNO}} (\nu)} \ -1 \ = \ 17.2\%$ \ \ \ \ \ \ (1)
\end{center}

It is important to examine if this discrepancy is significant as it could for 
example indicate (further) oscillation of (solar) electron neutrinos to sterile 
neutrinos. The uncertainties of the SSM prediction of 
solar neutrino fluxes were studied by Bahcall and 
Serenelli \cite{BS05} as listed in Table 1. 

\begin{table}[hb] 

\caption[TableI]{Fractional uncertainties of predicted \b8 and \xbe7 
solar neutrino fluxes \cite{BS05}}\label{tab1}
\    \\

\begin{tabular}{lllll}

\hline\\[-10pt]
Source \hspace{3cm} & \b8 \hspace{2cm} & \xbe7 \\
\hline\\[-10pt]
$pp (S_{11})$ & 1\% & 0.4\%  \\
$^3He+^3He (S_{33})$ & 2\% & 2\%  \\
$^3He+^4He (S_{34})$ & 8\% & 8\%  \\
$p+^7Be (S_{17})$ & 4\% & 0  \\
Composition (Z/X) & 12\% & 5\%  \\
Opacity $(\kappa)$ & 5\% & 3\%  \\
Diffusion & 4\% & 2\%  \\
Luminosity $(L_\odot)$ & 3\% & 1\% \\
\hline\\[-10pt]
\underline{Total:} & 16.7\% & 10.4\%
{\phantom{$00$}}\\
\hline
\end{tabular}
\end{table}

Currently the largest fractional uncertainty of the \b8 solar neutrino flux 
is due to the chemical composition (Z/X) which
is a consequence of the new evaluation of the abundance of heavy elements 
(mainly oxygen and neon) on the surface of the sun. The new evaluation leads to
doubling the uncertainty (from 6\% to 12\%) and to what 
was recognized as a new "standard solar model problem" vis-a-vis the predicted 
convective zone \cite{BSP}, and it was shown 
to be in conflict with SSM predictions based on helioseismology \cite{Basu}. 
Hopefully this large uncertainty due to Z/X will be resolved soon, 
and there is progress in reducing 
the error due to $S_{34}(0)$, see Table, leading to a considerably 
smaller total error. But in the same time the errors of nuclear inputs are far from 
resolved. In this paper we address the error of a crucial 
nuclear input, the formation of solar \b8 via the \be7pg reaction and its associated 
astrophysical cross section factor at zero energy, \S17.

The value of \S17 = 21.4 eV-b and the (impressive) accuracy of \xpm 4\% 
listed in Table 1 is adopted from the Seattle group \cite{Seatt}. It should be 
compared to the previously quoted value of \S17 = 19 +4 -2 eV-b \cite{Adel}.
The new quoted value of \S17 leads to an increase of 12.6\% in the predicted 
\b8 solar neutrino flux. 
We examine in details the claims of accuracy as well as the quoted value.
We demonstrate that the "world average" value of \S17 is driven by the Seattle result 
with an over estimated accuracy, and more realistic error bars are proposed. 
Due to large discrepancy between measured slopes of (mostly DC) data, and since theories 
also disagree on the predicted slopes, the value of \S17 is dependent on the choice of 
data and theory that one uses to extrapolate \S17. This situation must 
be alleviated by new measurement(s).

\subsection{Direct Capture}

\begin{figure}
 \includegraphics[width=6in]{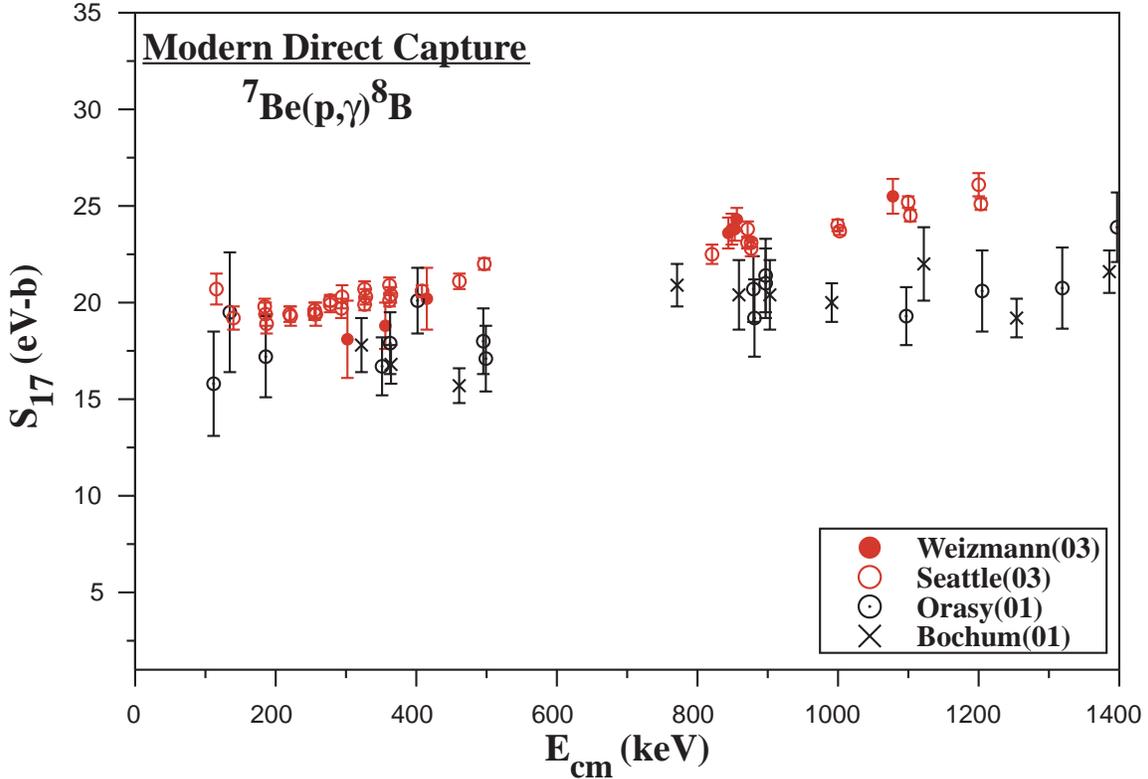}
 \caption{\label{World} Modern Direct Capture world data on \xs17 (excluding data 
around the 632 keV resonance) exhibiting marked systematical deviations 
by up to five sigma.}
\end{figure}

It is well known that the "old" data on Direct Capture (DC) \be7pg reaction 
\cite{Fil83,Vaughn,Parker,Kav} exhibit major systematic disagreements. But 
the situation is not improved with "modern" data on DC
\cite{Seatt,Ham01,Str01,Weiz}, as we show in Fig. 1. The data of the Orsay 
group \cite{Ham01} and Bochum group \cite{Str01} do not agree with that of the 
Seattle group \cite{Seatt} and Weizmann group \cite{Weiz}. As can be seen 
from Fig. 1 the disagreements among individual \xs17 data points are by as much as 
five sigma and there is not a single measured data point of the Bochum group 
that agrees with a data point measured by the Seattle group. Most disturbing is the 
systematic disagreement at low energies of the two "modern" high precision 
measurements of the Seattle \cite{Seatt} and Weizmann \cite{Weiz} groups.
These two measurement are in good agreement at higher energies (around 
1 MeV), but the Weizmann data are systematically below the Seattle data at 
low energies (300 - 500 keV), indicating a different slope.

The large discrepancies between measured individual DC data points shown 
in Fig. 1, demonstrate large systematical differences between 
"modern" DC data. These large systematical differences must be resolved 
before these data are used to extract \S17 from DC world data. The 
systematically disagreeing DC data can not be handled algebraically using 
statistical methods to extract average properties of DC data and in this 
case chi-square tests are meaningless. 

\section{The Slope Measured Between 300 - 1,400 KeV}

The slope of the measured astrophysical cross section factor between 300 
and 1400 keV is essential for extracting the d-wave contribution to measured 
data. This contribution must be subtracted from the data in order to 
extrapolate to zero energy \cite{Ham} where the s-wave dominate (approximately 
95\%). But so far we can only rely on theoretical estimates of the d-wave 
component \cite{Ham,DB,Jen,D} and these models can be tested only by a precise 
measurement of the slope.

The Seattle group extracted the average slope of DC data with 
high precision (\xpm 4.5\%), as shown in Fig. 19 of their paper \cite{Seatt}. But 
three of the five DC data points selected to be shown in Fig. 19 \cite{Seatt} 
(note that not all available DC data are shown in Fig. 19) have central values 
that disagree by a factor of 2.5 and almost three sigma. Furthermore, the quoted 
high precision is in fact driven by the claimed high precision of Seattle experiment 
hence in the next section we examine this precision in details.

\begin{figure}
\includegraphics[width=6in]{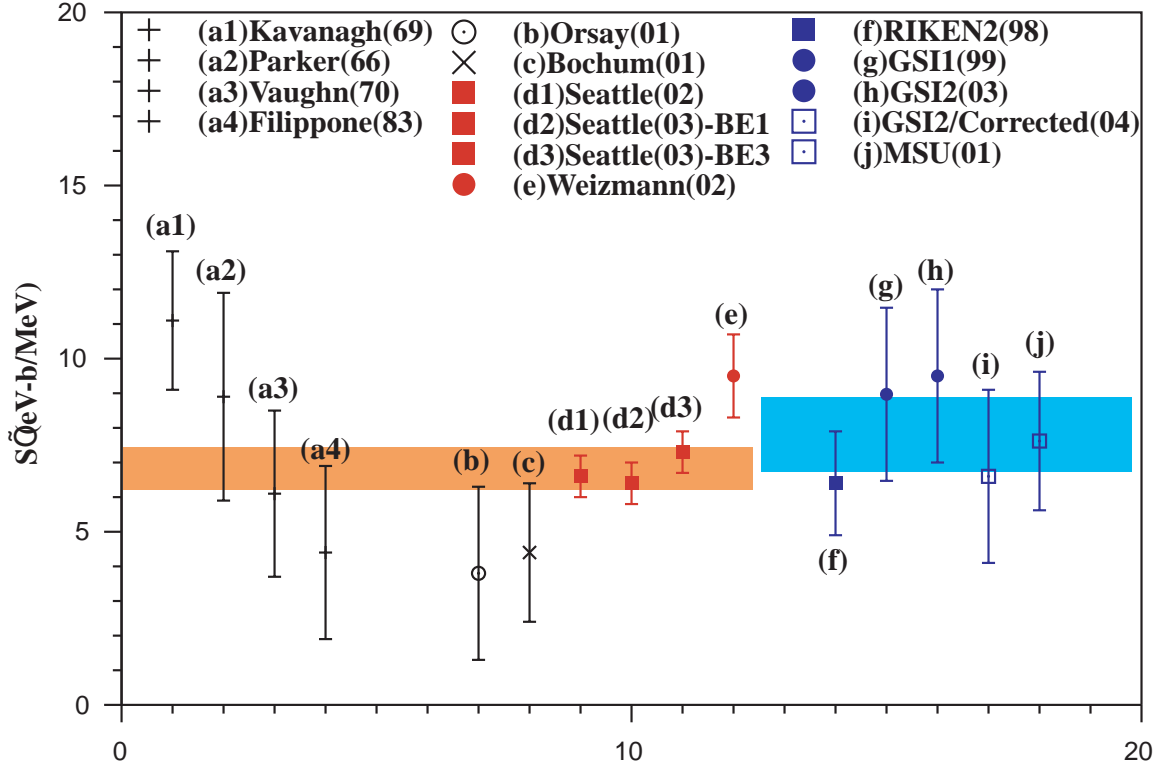}
 \caption{\label{SlopeS} The slopes (S' = dS/dE) 
of world data measured between 300 and 1400 keV. All available 
DC data are shown and the slope of the RIKEN2 data \cite{Kik} is shown 
correctly. The shown range of so called "average values" for DC and CD are  
in agreement as discussed in the text.}
\end{figure} 

In Fig. 2 we show the slope of all available DC data measured between 
300 - 1400 keV, including the data of 
Vaughn \cite{Vaughn}, Parker \cite{Parker}, and Kavanagh \cite{Kav} that 
were omitted in Fig. 19 of \cite{Seatt}. We also show all thus far measured 
CD data \cite{Kik,Iw99,Dav01,Sch03}.We conclude that the slope parameter 
can not be extracted from DC data with the (impressive) accuracy of 4.5\% 
\cite{Seatt},  unless one excludes some of the DC 
 measurements discussed above. An error which is at least a factor of 
 2 larger seems like a more reasonable choice. In the same time it is obvious 
 from Fig. 2 that there is no evidence for large disagreement of the slopes of CD 
 and DC data as claimed in \cite{Seatt}. In any case the large dispersion 
 of measured slopes shown in Fig. 2 can not allow us to determine the slope from 
 either CD or DC data with sufficient precision and this question must be addressed 
 by future experiments. Lack of accurate determination of the slope must lead to an 
 additional uncertainty due to the ill defined subtraction of the d-wave and the 
 extrapolation to zero energy.

\section{Accuracy of the Seattle Data}

Since the claimed high precision of the Seattle measured 
data points and quoted extracted \S17 \cite{Seatt} 
drive the value and the error of so called "world average" it seems reasonable to examine 
it in detail, especially at low energies. In Fig. 3 we show the results of the target-beam 
calibration data shown by the Seattle group using the $^7Be(\alpha,\gamma)^{11}C$ 
reaction. As can be seen in Fig. 3 both calibration 
spectra published in the Phys. Rev. Lett. and the Journal Phys. Rev. 
C (which where measured with different \xbe7 targets) exhibit a resonance energy 
which is off by 9 keV from the known energy in $^{11}C$. This 9 keV shift is intolerable and 
the authors as yet did not explain this discrepancy in 
a published erratum. The repeated publication of 
the same mistake first in the PRL paper and a year later in the PRC long paper does 
not lend credence to the claimed high precision experiment.

\begin{figure}
\includegraphics[width=5in]{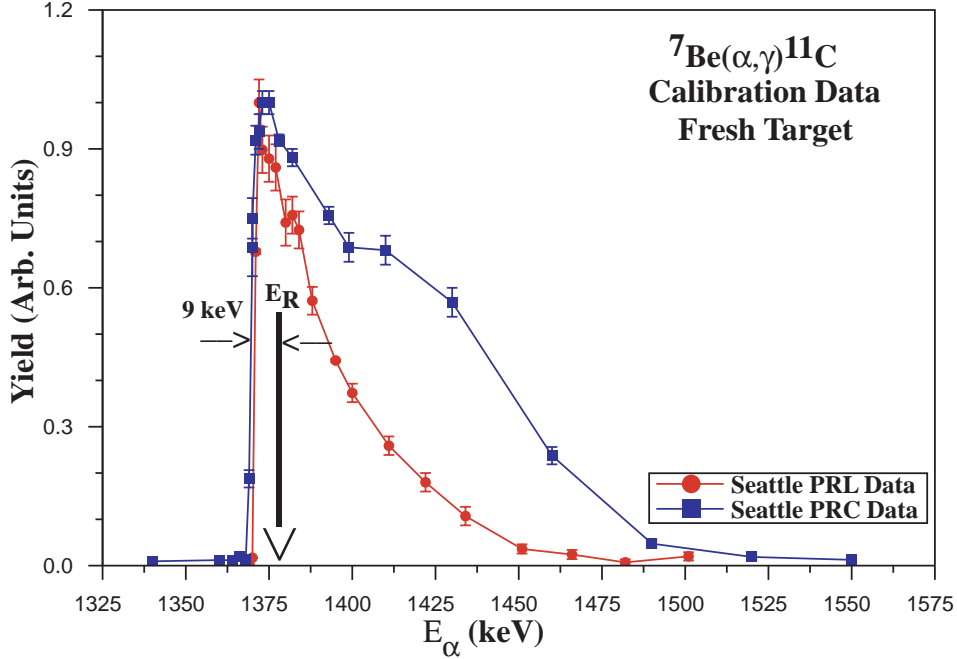}
 \caption{\label{beam} The measured beam-target calibration spectra \cite{Seatt} using 
the $^7Be(\alpha,\gamma)^{11}C$ reaction. Both spectra published in PRL and PRC 
exhibit resonance energy which is off by 9 keV from the well known resonance energy 
in $^{11}C$ as indicated.}
\end{figure} 

Furthermore, in their paper they show the \xbe7 target profile measured in the middle 
 (PF2) and at the end (PF3) of the experiment \cite{Seatt}. 
 These profiles are superimposed on top 
of each other in Fig. 4a, from which it becomes clear that we can not support the claim 
\cite{Seatt} that "they are similar". In fact there is evidence that the \xbe7 moved further 
into the target in the intervening period between these two measurements. The 
unstable nature of the \xbe7 during the experiment must be reflected in a systematic 
uncertainty due to variation of the expected yield, as we discuss below. Another systematic 
uncertainty that is not considered here is associated with the evaluation of the 
effective center of mass energy. 

\begin{figure}
\includegraphics[width=4in]{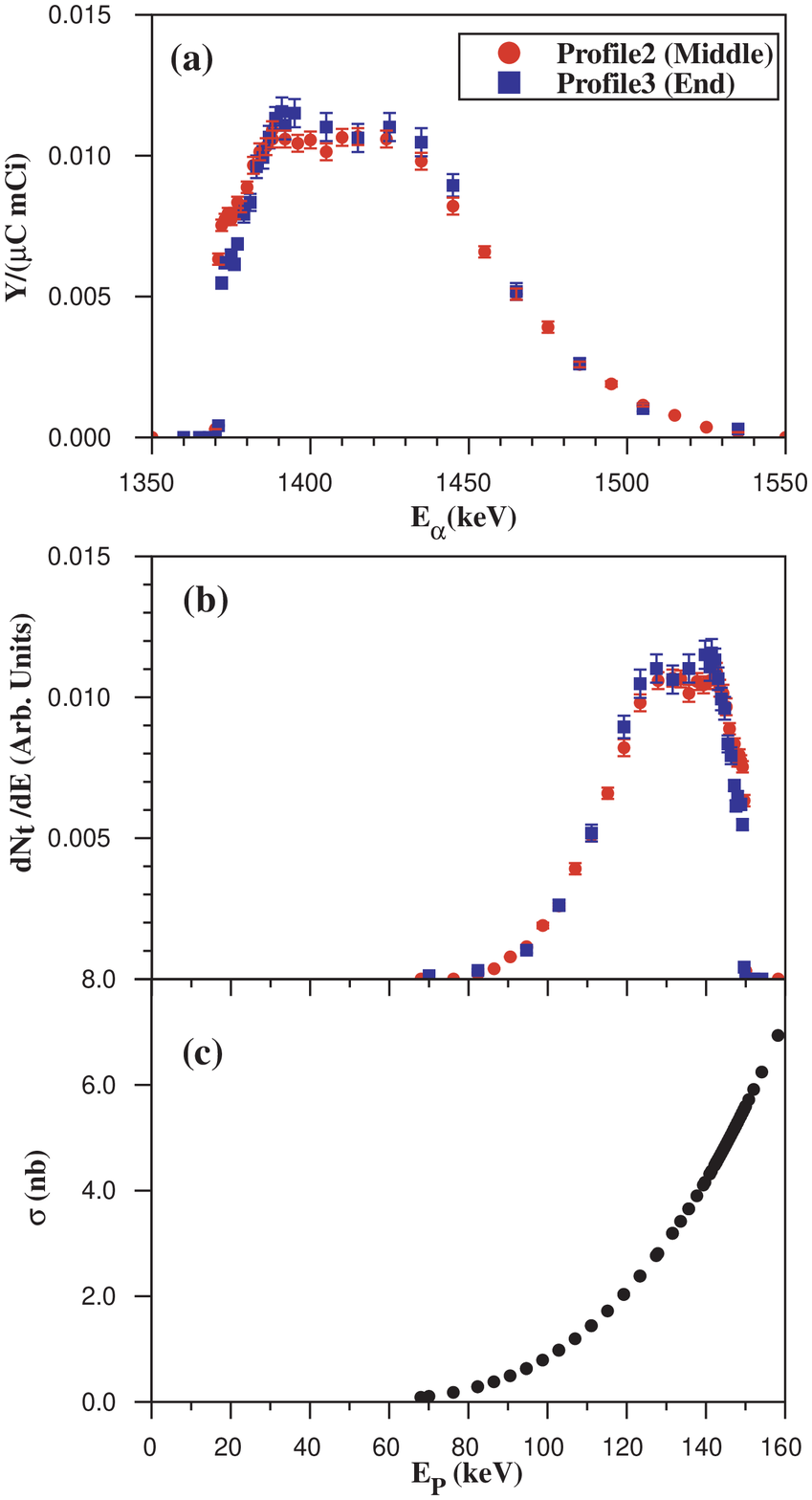}
 \caption{\label{target} (a) Target profiles measured using 
the $^7Be(\alpha,\gamma)^{11}C$ reaction in the middle (PF2) and 
at the end (PF3) of the experiment \cite{Seatt}, (b) the target profile 
for a 150 keV proton beam, (c) the predicted cross section variation across 
the target.}
\end{figure} 

These profiles are measured with alpha-particle beams 
with energies of approximately 1.4 MeV in steps 
of $\Delta$E = E$_\alpha$ - E$_R$ which can be translated to steps of 
 $\Delta$E = (E$_\alpha$ - E$_R)\times 
{dE \over dX}_{p + (Mo + ^7Be)}$/${dE \over dX}_{\alpha + (Mo + ^7Be)}$ for 
 proton beams. Note that the ratio 
 ${dE \over dX}_{p + (Mo + ^7Be)}$/${dE \over dX}_{\alpha + (Mo + ^7Be)}$
 is almost independent of the exact $Mo \ + \ ^7Be$ mixture of the target \cite{Seatt}, 
 as well as variation  among the various tabulated individual energy loss.
 Hence for an incoming proton beam (E$_p$) the
\xbe7 target profiles during the \be7pg reaction measurement is given by: 

\begin{center}
 ${dN_{tgt} \over dE}((E_p \ - \ (E_\alpha -1370) 
 \times {{dE \over dX}_{p + (Mo + ^7Be)} \over {dE \over dX}_{\alpha + (Mo + ^7Be)}}) \ \ 
 \propto$ \ Profile$(E_\alpha)$ \ \ \ \ \ \ (equ. 2)

\end{center}

\begin{figure}
\includegraphics[width=5in]{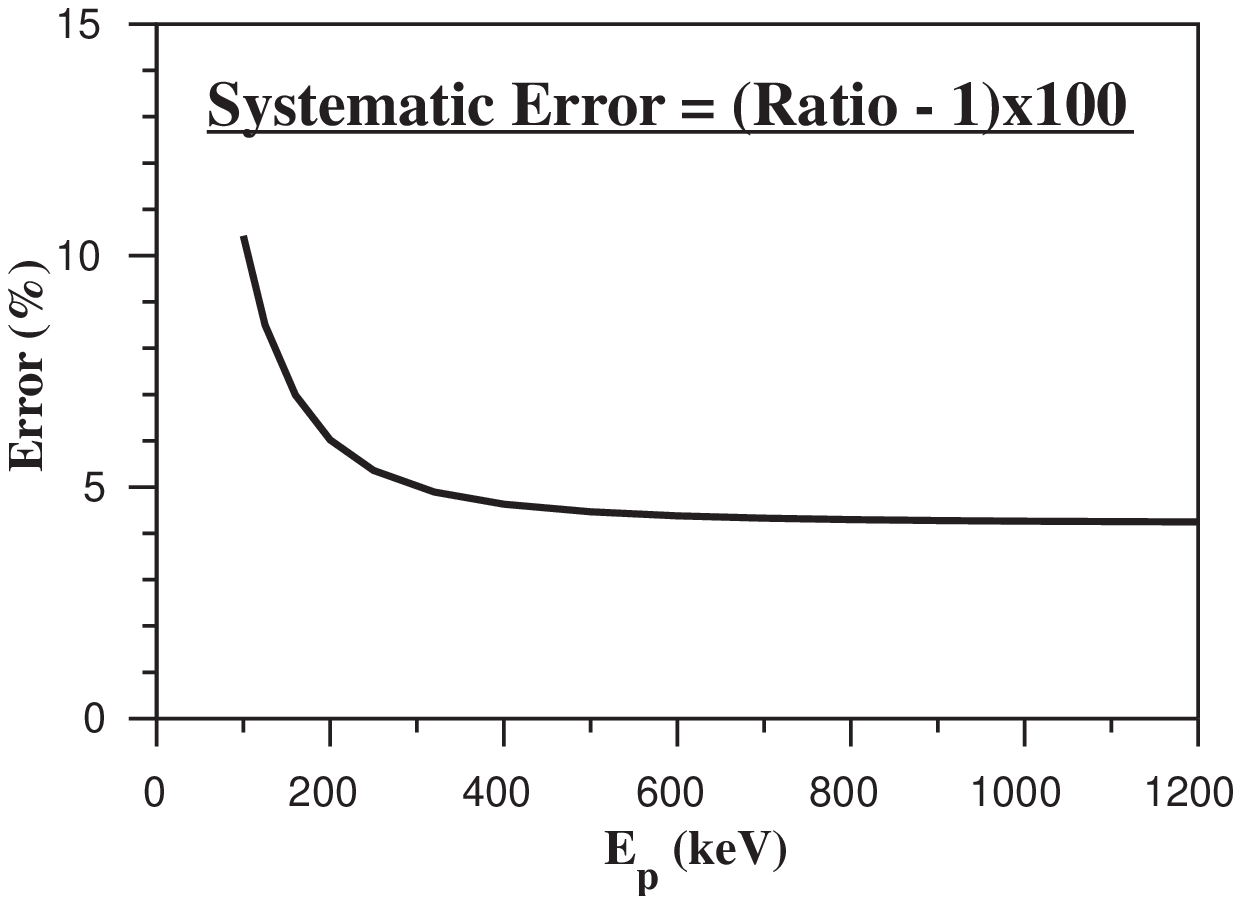}
 \caption{\label{error} One systematic error of the Seattle experiment \cite{Seatt} 
due to variation in \xbe7 target profile shown in Fig. 4. The two profiles measured 
in the middle (PF2) and at the end (PF3) of the experiment are used.}
\end{figure} 

\begin{figure}
\includegraphics[width=5in]{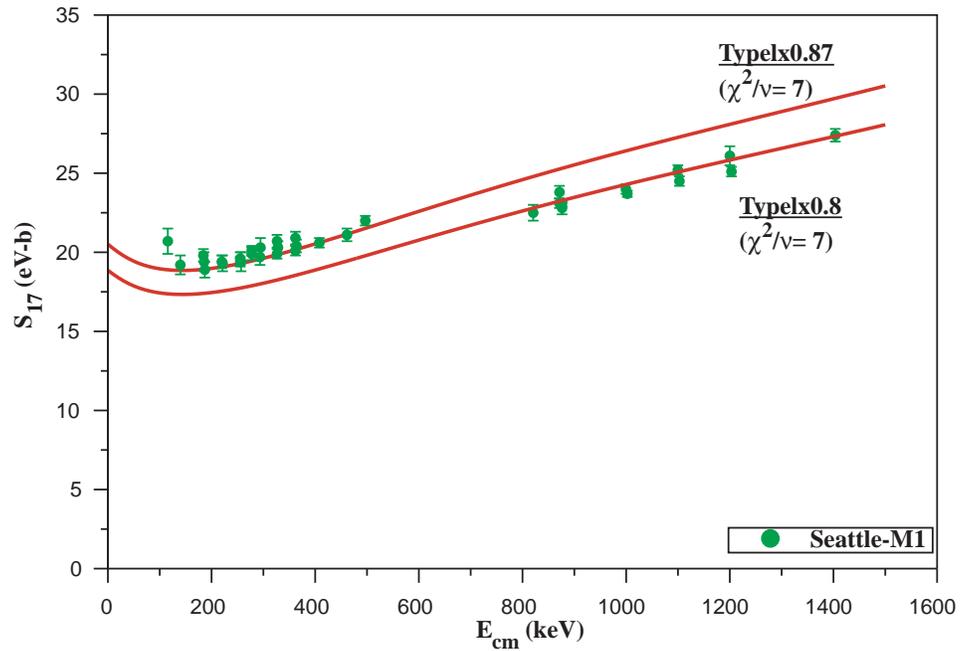}
 \caption{\label{Type} A comparison of the Typel potential model \cite{Sch03} 
with the Seattle data \cite{Seatt}.}
\end{figure} 

\begin{figure}
\includegraphics[width=5in]{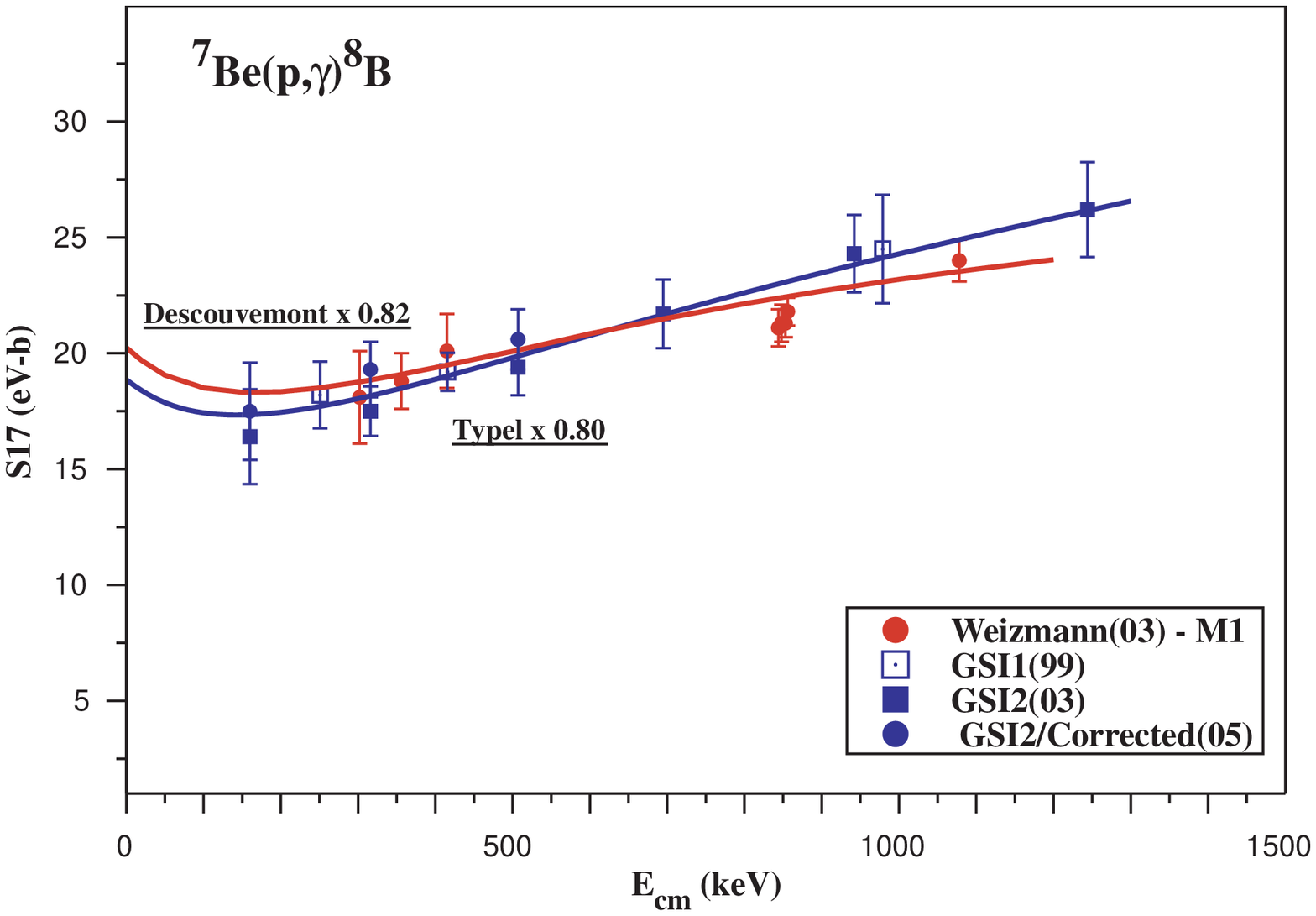}
 \caption{\label{Theory} A comparison of the Descouvemont cluster model 
\cite{D} and Typel potential model \cite{Sch03} with the GSI \cite{Sch03} 
and Weizmann data \cite{Weiz}.}
\end{figure} 

The so obtained profile is shown in Fig. 4b, it leads to very large target thickness,  
for example of the order of 35 keV at 
a beam energy of approximately 130 keV. Over such a thick target the cross section 
varies by almost a factor of five, as shown in Fig. 4c. A convolution of the so 
obtained target profile with the expected variation  
of the cross section leads to the expected yields:

\begin{center}

Yield  $\propto \ \ \Sigma_{E_i} {dN_{tgt} \over dE_p}(E_i) \times \sigma(E_i)$ 
\ \ \ \ \ \ \ \ \ \ \ \ \ \  (equ. 3)

\end{center}

The yields evaluated for the two shown profiles measured in the middle (PF2) and at the 
end (PF3) of the experiments differ by 7.5\% at the lowest measured energy, 
almost a factor of 3 larger than the quoted systematic uncertainty \cite{Seatt}. In Fig. 5 
we show the differential yields = Y(PF2)/Y(PF3) - 1 evaluated for all incoming 
proton beam energies. This differential yield must be 
considered a lower limit of the systematical uncertainty. This uncertainty is a 
factor of 3 larger than the quoted varying systematical uncertainty \cite{Seatt}.
Clearly the most troubling uncertainty is at energies below 400 keV. This energy 
range was emphasized in the Seattle paper as most useful for extracting \S17, 
but it is measured with the largest systematical uncertainty.

The large systematical uncertainty of Seattle data at low energies is particularly 
bothersome vis-a-vis the exclusion of theoretical models such as the potential 
model of Typel \cite{Sch03}. In Fig. 6 we show a 
comparison of that model to the Seattle data 
where we find that the model reproduce well the high energy data but not the 
low energy data (or vice versa). The large systematical error at low energies 
shown in Fig. 5 cast doubt on the claim that the Typel model is "rejected" by the 
data. This model fits well all other data and consistently yield extrapolated 
values of \S17 which are approximately 3 eV-b smaller, as shown in Fig. 7.

We conclude that in the absence of an accurate measurement of the slope we 
must add an extrapolation error to the quoted \S17. And 
in view of the fact that the Typel model is consistent with all existing data except 
the Seattle data, and in view of the large systematical error of the Seattle data 
at low energies, a reasonable estimate of an extrapolation error is of +0.0 
-3.0 eV-b. Such an a large error implies the discrepancy of the SSM shown in equ. 1 
is not significant solely due to nuclear uncertainty even if the uncertainty due to chemical 
composition will be reduced.


\end{document}